\documentclass[9pt,journal]{IEEEtran}
\usepackage{amsmath,amsfonts}
\usepackage{algorithmic}
\usepackage{algorithm}
\usepackage{array}
\usepackage[caption=false,font=normalsize,labelfont=sf,textfont=sf]{subfig}
\usepackage{textcomp}
\usepackage{stfloats}
\usepackage{url}
\usepackage{verbatim}
\usepackage{graphicx}
\usepackage{amssymb}
\usepackage{color}
\usepackage{multirow}
\usepackage{booktabs}
\usepackage{tabularx}
\usepackage{enumitem}
\usepackage{cite}
\usepackage{hyperref}

\usepackage{newtxtext}

\usepackage[normalem]{ulem}
\DeclareMathOperator*{\argmin}{arg\,min}

\begin{document}

\title{Structure-Preserving Medical Image Generation from a Latent Graph Representation}

\author{Kevin Arias,~\IEEEmembership{Student member,~IEEE,}
Edwin Vargas,~\IEEEmembership{Member,~IEEE,}
Kumar Vijay Mishra,~\IEEEmembership{Senior Member,~IEEE,}\\
Antonio Ortega,~\IEEEmembership{Fellow,~IEEE,}
Henry Arguello,~\IEEEmembership{Senior Member,~IEEE,}
\thanks{K. A., and H. A. are with the CS Department, Universidad Industrial de Santander, Bucaramanga 680002 Colombia,} 
\thanks{E. V is with the ECE Department,  Rice University, TX 77005 USA,}
\thanks{K. V. M. is with the United States CCDC Army Research Laboratory, Adelphi, MD 20783 USA,}
\thanks{A. O is with the ECE Department, University of Southern California, CA 90089, USA.}
}

\maketitle

\begin{abstract}
Supervised learning techniques have proven their efficacy in many applications with abundant data. 
However, applying these methods to medical imaging is challenging due to the scarcity of data, given the high acquisition costs and intricate data characteristics of those images, thereby limiting the full potential of deep neural networks. 
To address the lack of data, augmentation techniques have been explored, leveraging geometry, color, and the synthesis ability of generative models (GMs). Despite previous efforts, gaps in the generation process limit the impact of data augmentation to improve understanding of medical images, e.g., the highly structured nature of some domains, such as lung X-ray images, is ignored. Current GMs rely solely on the network's capacity to blindly synthesize augmentations that preserve semantic relationships of chest X-ray images, such as anatomical restrictions, representative structures, or structural similarities consistent across datasets. 
In this paper, we introduce a novel generative framework that leverages the structural resemblance of medical images by learning a latent graph representation (LGR). 
We design an end-to-end model to learn (i) a LGR that captures the intrinsic structure of lung X-ray images and (ii) a graph convolutional network (GCN) that reconstructs the lung X-ray image from the LGR. 
We employ adversarial training to guide the generator and discriminator models in learning the distribution of the learned LGR. Using the learned GCN, our approach generates structure-preserving synthetic images by mapping generated LGRs to lung X-ray. Additionally, we evaluate the learned graph representation for other tasks, such as X-ray image classification and segmentation.
Numerical experiments demonstrate the efficacy of our graph in capturing semantic relationships that enhance lung X-ray augmentation with a performance increase of up to $3\%$ and $2\%$ for classification and segmentation, respectively.
\end{abstract}


\section{Introduction}
\IEEEPARstart{T}{he} continuous advancements in deep learning have demonstrated its potential for analysis and diagnosis in medical imaging. 
In particular, deep network architectures (DNA) \cite{anaya2021overview}, such as convolutional neural networks (CNNs), recurrent neural networks, autoencoders, attention models, transformers, and graph convolutional networks (GCNs) 
have been employed for classification of pathology \cite{jiang2023review, yang2023medmnist}, organ and anomaly segmentation \cite{wang2022medical, luo2022semi}, detection of abnormalities \cite{jiang2023review, tsuneki2022deep} and anomaly localization tasks \cite{abdou2022literature}. 
Training pipelines should help the DNAs understand the intricate characteristics inside medical datasets, e.g., highly structured images and anatomical relationships preserved along the dataset.
For medical diagnosis with limited training data, state-of-the-art (SOTA) DNAs are trained using neural network backbones pre-trained with larger datasets, e.g., natural image datasets. 
These approaches often have significant performance limitations because pre-training data is not directly relevant to the target task. 
Most approaches overcome the limitations of DNAs in scarce medical data contexts by focusing on (i) leveraging data augmentation techniques to increase generalization capabilities and (ii) incorporating specific prior knowledge into the network architecture design or the training process.

Medical data augmentations have been proposed to increase the number of images used for training. These augmentations are derived from geometric transformations, color space transformation, or noise injection \cite{shorten2019survey, goceri2023medical}, designed to build invariance \cite{cosentino2022geometry}. 
However, the resulting augmented images are highly correlated with the original samples, making it hard to increase generalization capabilities, as required to improve network performance \cite{garcea2022data}.
Increasing generalization may require increasing augmentation strength (e.g., applying noise with higher variance). However, this limits how well the semantic meaning, e.g., anatomical details, is preserved, raising questions among medical experts about image realism and their utility for training models that support decision-making. 
As an alternative, significant recent advances have been made in augmentations using generative models (GMs), 
which can produce images with greater realism and variability \cite{kebaili2023deep}. 
Conventionally, GM architectures such as variational autoencoders (VAEs) \cite{ehrhardt2022autoencoders}, generative adversarial networks (GANs) \cite{singh2021medical}, and diffusion models (DMs) \cite{kazerouni2022diffusion} are trained using 2D medical images to capture their pixel-level distribution. 
Generated images following the learned distribution have been used as augmentations to feed networks, increasing their generalization and performance in medical analysis and diagnosis tasks \cite{kebaili2023deep, chlap2021review, goceri2023medical}.
In particular, GMs have been employed to generate effective augmentations for classification \cite{pesteie2019adaptive,zhuang2019fmri,chadebec2022data} and segmentation \cite{pesteie2019adaptive, huo2022brain} tasks on brain, lung, breast, and eye images acquired from different imaging modalities \cite{goceri2023medical}. 
Some of these generative architectures have also been adapted by SOTA methods to augment chest X-ray images. 
However, images from GNs tend to often resemble only a limited subset of the target distribution \cite{bau2019seeing} \cite{srivastava2017veegan}, which limits the ability to increase the generalization capabilities of the network. 
Although our ideas are generalizable to a range of medical imaging tasks, in what follows, we focus on chest X-ray images.

Our work is motivated by noting that existing GMs do not integrate structural knowledge into their augmentations. Structural priors lead to significant performance improvements in segmentation or classification for highly structured medical images.
For example, incorporating information on anatomical structures into the model training leads to improvements in segmentation \cite{hosseinzadeh2023towards}. 
Closer to our application domain, recent models that process chest X-ray images towards a COVID-19 diagnosis have relied on the design of graph-based networks, assuming that structural relationships between regions in chest X-ray can be well-captured by graph connectivity \cite{tang2022nscgcn}. 
Such improvements in precision or accuracy can be achieved even when the structural priors are captured by a simple graph construction limited to an $8$-neighborhood local topology \cite{tang2022nscgcn}, owing to the similarity in image acquisition for a particular imaging modality.
For example, chest X-rays exhibit similar structural features, such as the positioning of the lungs, internal organ distributions, and bone structures across all the images acquired within this modality. 
Our key observation is that the underlying structure of chest X-ray images captured by graphs has not yet been leveraged for data augmentation. 
Instead, current GMs rely on complex network architectures, high-computing training processes, prior knowledge, additional data, and difficult-to-set-up learning pipelines to capture the underlying structure of medical data and compensate for the lack of training data. 
   
\subsection{Prior art}

\noindent \textbf{Transformations on the original images}. 
SOTA works have employed different transformations as data augmentation for different medical imaging modalities, such as brain MRI \cite{liu2022cerebrovascular}, CT \cite{tandon2022vcnet}, X-ray \cite{woan2022multiclass}, or retinal imaging. These transformations include rotation, rescaling, shearing, flipping, shifting, cropping, zooming and brightness and have been used to augment the original dataset for tumor classification \cite{srinivas2022deep}, image classification \cite{haq2022iimfcbm}, vessel segmentation \cite{liu2022cerebrovascular}, COVID-$19$ detection \cite{woan2022multiclass}, lesion classification \cite{ueda2022development}, soft tissue classification\cite{sabani2022bi}; and glaucoma identification \cite{shyamalee2022cnn}. The main limitation of these transformations is that, while they aim to improve generalization, they often compromise the semantic integrity and realism of the medical images, which can negatively impact performance in medical diagnostic tasks. For example, adding high-variance noise or applying large-angle rotations to X-ray images can result in samples that no longer appear realistic. 

\noindent \textbf{Generation of artificial images}.
Generation-based traditional augmentation techniques create new realistic images that increase the diversity of transformation-based augmentations \cite{chen2022generative}. 
These augmentations are learned from GNs, such as VAEs and GANs, 
which can approximate the true data distribution from samples drawn from a random latent space. 
Then, learned GNs randomly sample random latent vectors and obtain new images to augment the datasets. 

For the GAN architecture, two models, the generator and discriminator, are trained adversarially to generate and qualify the realism of the images, resulting in a learned generator to create realistic images from random latent vectors. Different variants of GAN have been successfully employed for augmentation of brain MR images, lung CT images, mammography images, and eye fundus images, increasing the performance of a task such as vessel segmentation \cite{kossen2021synthesizing}, 
tumor segmentation \cite{sun2020mm}, anomaly classification \cite{jha2022framework, onishi2020multiplanar}, image classification \cite{desai2020breast, ju2021leveraging, toda2021synthetic} and lesion detection \cite{shen2021mass}. 

For the VAE architecture, encoder and decoder networks are sequentially trained to reconstruct the image, and the latent space is learned at an intermediate point after the encoder to estimate its posterior distribution \cite{kebaili2023deep}. Unlike GAN, the VAE is trained to maximize the likelihood of the data rather than adversarially. Although VAEs are better at approximating the distribution of real data, they have not been effectively used for medical image augmentation due to the blurry nature of the generated images. 
Alternatively, VAE-based medical image augmentation combines VAEs and GANs to exploit the advantages of both models. VAE-GAN architecture introduces the adversarial objective of GAN on the VAE objective to improve the generated images' sharpness while preserving the VAEs' ability to learn a compact latent space \cite{liang2021data, ahmad2022brain}. Also, conditional VAEs (CVAEs) have addressed the medical image augmentation when additional information is available, such that the generation is conditioned on additional information, e.g., class label or attributes, to more concisely represent specific subgroups \cite{zhuang2019fmri, biffi2018learning}. Even though images generated by GNs outperform images obtained from simple transformations, they often exhibit high similarity to specific subsets within the target data distribution, thereby constraining their effectiveness in enhancing the generalization capacity of the model.

\noindent \textbf{Diffusion Probabilistic models}.
Augmentation techniques based on diffusion models (DMs) have shown more realistic and high-quality image synthesis than GANs and VAEs. DMs are based on the diffusion process, which gradually adds Gaussian noise to the images of the distribution, while the diffusion inverse process, known as generation, is modeled as a denoising process. This strategy of modeling the target distribution from step-by-step simpler distributions has made it possible to model more complex structures. 
DMs for synthesizing lung X-ray and CT images have shown potential for medical image generation \cite{ali2022spot}. Other approaches for diffusion-based medical image augmentation present variations on the DM architectures, e.g., latent DMs (LDMs) work as a combination of autoencoders and DMs where the autoencoder is employed to map the image to a lower-dimensional latent representation \cite{pinaya2022brain}. Also, the combination of VAE-GANs and DMs has been used to generate images with the segmentation label, with the generative models VAE-GANs and DMs generating the segmentation map and image, respectively \cite{fernandez2022can}. Although 
conditional generation using DMs is promising for medical image segmentation, it remains a challenge when compared with \textit{only} GAN-based approaches because DMs require significantly more computation, with longer training and sampling times \cite{usman2024brain}.

\noindent \textbf{Construction of latent graphs for medical diagnosis}.
Recent approaches have explored the construction of graphs in the latent space to encode structural priors that enhance the representation of chest X-ray images. By modeling the relationships among latent features as graph connectivity, these methods aim to capture the underlying spatial and semantic correlations inherent in the imaging data. Authors in \cite{elazab2022novel} proposed a multi-site graph convolutional network with a supervision mechanism, where graphs are constructed in the latent feature space to integrate information across different sites. 
However, in the graphs constructed in \cite{elazab2022novel}, each image is represented by a node and a feature vector. In contrast, in our work, we are interested in intra-image spatial relationships. 
Similar to our idea, \cite{tang2022nscgcn} introduced the NSCGCN model, which forms a latent-space graph to exploit high-level feature interactions and guide the GCN learning process for more accurate classification.
However, \cite{tang2022nscgcn} adopts a patch-level graph construction that restricts the connectivity to a fixed 8-neighborhood local topology. In our work, we also learn a graph connecting patches, but unlike \cite{tang2022nscgcn}, we use no prior locality restrictions and learn the weights and the graph sparsity from data. 

\subsection{Our contributions}
\par\noindent
\textbf{Overall goal:} We aim to improve image augmentations for training, where the new images generated provide increased diversity, while remaining semantically meaningful. 
To understand the intuition behind our approach, assume that typical task-relevant images (e.g., chest X-ray images) belong to a low-dimensional manifold within the space of all images of the same size. In practice, it is hard to determine if generated images belong to the data manifold, but we can use proxies, such as comparing the Euclidean distance between images and their Fréchet inception distance (FID) \cite{yu2021frechet}, for this purpose. 
From this perspective, methods based on transformations do not restrict changes to the images to be aligned with the manifold. For example, adding noise to images as an augmentation will place the augmented images in a hypersphere centered on the original example, and they will show a large FID with respect to images in the dataset. Other existing methods, which rely on VAEs and GANs, can improve manifold ``alignment'' (i.e., keep FID lower) compared to transformation-based methods, at the cost of a small Euclidean distance.  
Our goal is to improve manifold alignment further while achieving large distances (i.e., increasing the Euclidean distance while maintaining the FID low). To this end, our main contributions are:
\par
\begin{figure*}
    \centering
    \includegraphics[width=0.96\linewidth]{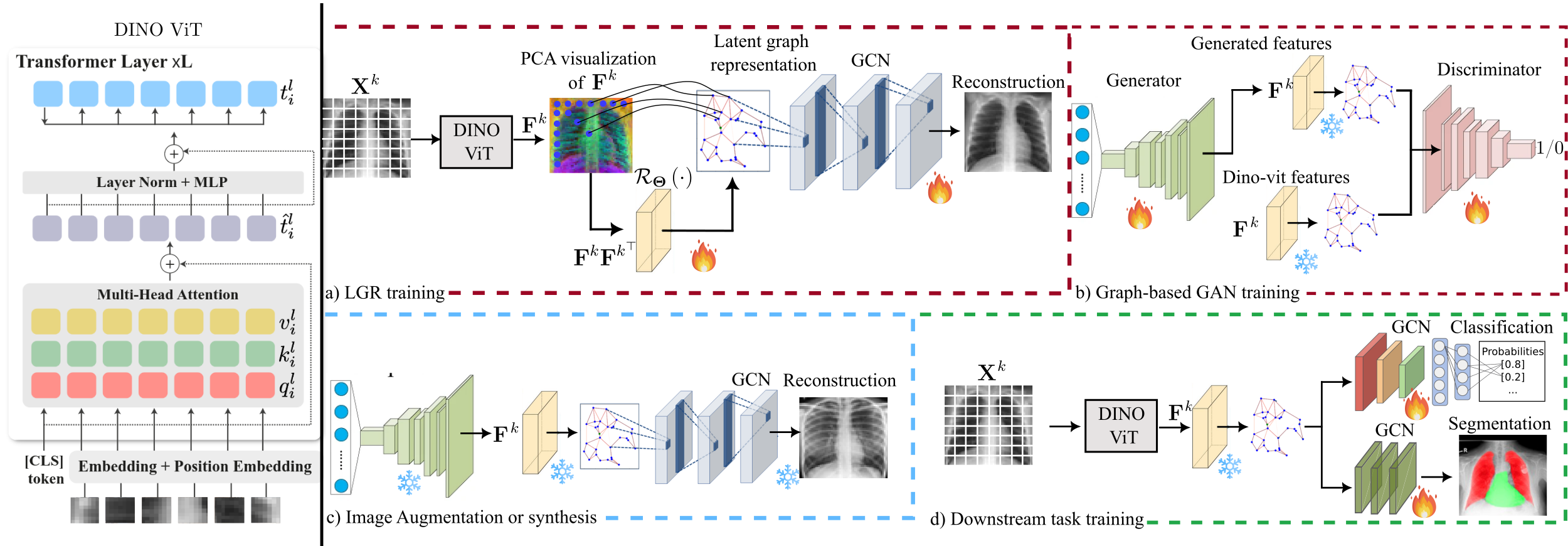}
    \caption{Overview of the proposed structure-preserving image analysis. Modules annotated with fire symbols are trainable, and those annotated with snowflake symbols are frozen. a) LGR training: A latent graph representation is end-to-end learned to capture the structural resemblance of the chest X-ray images. b) Graph-based GAN training: 
    Adversarial learning is performed to generate new graph representations; the high potential of GANs is employed to learn the graph distribution. c) Image augmentation or synthesis: on inference, new realizations of random noise can generate new graphs that are passed through the GCN to create structure-preserving images. d) Downstream task training: Chest X-ray images are first represented using our LGR. The resulting graphs are then used to optimize a GCN for the classification or segmentation tasks.}
    \label{fig:intro_fig}
\end{figure*}

\noindent
\textbf{1) Novel latent graph representation (LGR) construction.}
We propose a novel LGR construction to capture the semantic and structural relationships of chest X-ray images. Specifically, the latent graph (see \autoref{fig:intro_fig}(a)) is such that (i) the signals in each vertex are vision transformer (ViT) features from corresponding spatial image patches, selected to exploit their semantic richness \cite{amir2021deep}, 
(ii) the strength of the connections between features is calculated from a correlation measure, and 
(iii) the topology is learned from a CNN $\mathcal{R}_{\boldsymbol{\Theta}}$ with binary output to select or remove the connections between the vertices. 
We propose self-supervised learning of the graph topology by jointly optimizing $\mathcal{R}_\Theta$ and a structured mapping that produces images from the LGR. Thus, the structure of the graph depends on the learned patch relationships within the input image. 

\par\noindent
\textbf{2) Harnessing structural properties of images for GM.}
We leverage our proposed LGR in the adversarial training of a GM (see \autoref{fig:intro_fig}(b)) that can generate semantic LGRs that preserve the structural properties of the chest X-ray database. Then, we generate structure-preserving images by feeding the generated LGRs to the learned graph-to-image mapping as shown in \autoref{fig:intro_fig}(c). 
Our approach achieves a 30-point reduction in the FID metric compared to baseline GANs, showing that the generated images remain close to the task-related data manifold. The Euclidean distance between our generated images and the dataset images is greater than that of competitive GANs, demonstrating that we can enhance diversity (larger Euclidean distance) without compromising the fidelity of the distribution (lower FID).
We demonstrate that the proposed generated images enhance data augmentation of chest X-ray images for both classification and segmentation tasks. Specifically, feeding baseline networks with our proposed structured augmentations results in improvements of up to $3 \%$ in classification accuracy and $2 \%$ in the DICE metric for segmentation.  \\
\textbf{3) LGR as a standalone encoding for GCNs in medical image analysis.} 
Our proposed LGR construction can also serve as a graph representation method of X-ray images (see \autoref{fig:intro_fig}(d)) in the pneumonia classification and lung segmentation tasks using GCN, 
where, instead of learning the graph topology using a self-supervised approach, we directly optimize the proposed LGR with a GCN that performs pneumonia classification or lung segmentation. 
The results show that our LGR improves pneumonia classification accuracy with GCN by $1.25 \%$, demonstrating a higher representational power than other SOTA representations for GCNs. 

\textbf{Notation:} This paper uses boldface lowercase and uppercase for vectors and matrices, respectively. The $i$-th entry of the vector $\mathbf{x}$ is $\mathbf{x}_i$. $\mathbf{X}_{i,j}$ represents the $(i,j)$-th entry of matrix $\mathbf{X}$. $\mathbf{X}_{i}$ denotes the $i$-th column of $\mathbf{X}$. We denote the transpose, conjugate, and Hermitian of a matrix by $\mathbf{X}^\top$, $\mathbf{X}^*$, and $\mathbf{X}^H$, respectively. Sets, functions, or graphs are represented using calligraphic letters. $\circ$ is the entrywise product. $\operatorname{vec}(\cdot)$ is the vectorization operator that transforms a matrix into a column vector by stacking its columns.

\section{Problem Formulation}

Consider X-ray images of a given dataset $\mathcal{X} = \left\lbrace\mathbf{X}^{k} \right\rbrace_{k=1}^{K}$, such that $\mathbf{X}^{k} \in \mathbb{R}^{M \times N}$ and $M \times N$ size. To build our specialized graph structure, we divide a given image $\mathbf{X}$ into non-overlapping patches of size $P \times P$. Then, we define an undirected weighted graph  $\mathcal{P} = \left( \mathcal{V}, \mathcal{E}, \mathbf{W} \right)$ containing a set $\mathcal{V}$ with $ V = NM / P^{2}$ vertices corresponding to the image patches and a set $\mathcal{E}$ of edges. Each edge is undirected and is given an edge weight $\mathbf{W}_{i,j}$ that represents the similarity between patches $i$ and $j$, with $1$ corresponding to the maximum similarity. 
We propose constructing the edge matrix leveraging vision transformer (ViT) features. Specifically, we define a high-dimensional signal residing on the vertices of this graph ~$\mathcal{T}: ~\mathcal{V} \rightarrow ~\mathbb{R}^D$  as the $D$-dimensional \textit{key} token of ViT features. The matrix representation of this signal is $\mathbf{F} = \left[\mathbf{f}_1, \cdots,  \mathbf{f}_V\right]^\top$ $\in \mathbb{R}^{V \times D}$, where $\mathbf{f}_i$ is the signal value at vertex (patch) $i \in V$. Based on this representation, we define the edge weights as:

\begin{equation}
    \mathbf{W} = \mathcal{R}_{\boldsymbol{\Theta}} \left( \mathbf{C} \right) \circ \mathbf{C},
\label{eq:weight_matrix}
\end{equation}
where $\mathbf{C} = \mathcal{N} \left( \mathbf{F} \mathbf{F}^{{\top}} \right)$ is a normalized correlation matrix of the ViT features, $\mathcal{N} \left( \cdot \right)$ is a function that produces outputs in $[0,1]$, and  $\mathcal{R}_{\boldsymbol{\Theta}} \left( \cdot \right)$ is a convolutional network with learnable parameters $\boldsymbol{\Theta}$. 
The output of $\mathcal{R}_{\boldsymbol{\Theta}} \left( \cdot \right)$ is a sparse binary selection matrix, with the same dimensions as the input correlation matrix, that defines the structure of the graph.
Henceforth, we denote $\mathcal{P}^k = ~\left( \mathbf{F}^k, \mathbf{W}^k \right)$ as the LGR of a given sample X-ray image $\mathbf{X}^{k} \in \mathcal{X}$, where $\mathbf{F}^k$ are its ViT features, and $\mathbf{W}^k$ its corresponding weights computed using \eqref{eq:weight_matrix}.

\begin{figure}[t]
	\centering
	\includegraphics[width=\linewidth]{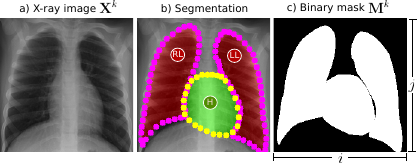}
	\vspace{-10pt}
	\caption{Foreground and background segmentation: anatomical segmentation is performed using \cite{gaggion2022improving} as a middle stage to extract the regions of the right lung (RL), left lung (LL), and heart (H). Then, we construct a binary mask $\mathbf{M}^{k}$ that discriminates between background pixels and pixels inside the lung and heart, named foreground. Foreground pixels will be given priority for the graph construction since these are the pixels involved in giving a verdict in the pneumonia classification task.}
	\label{fig:segmentation}
\end{figure}

In this work, we aim to learn the LGR of chest X-ray images, i.e., learn $\boldsymbol{\Theta}$ to capture their structural relationships. To achieve this goal, we propose a self-supervised approach that jointly optimizes the parameters $\boldsymbol{\Theta}$ to define the graph topology and a structured inverse mapping that produces images from graphs. Hence, we leverage structural dependencies within the data and do not require labeled data.
Furthermore, we also propose differentiating between foreground and background correlations to emphasize the foreground, which is the region of interest for medical diagnosis. 
More precisely, consider the binary mask obtained using \cite{gaggion2022improving} (see \autoref{fig:segmentation}(c)) segmenting the foreground $\left( \mathbf{M}^{k} = 1 \right)$ and background $\left( \mathbf{M}^{k} = 0 \right)$. Then, the foreground correlation values are 
$\mathbf{\dot C}^{k}_{i, j} = \mathbf{C}^{k}_{i, j}$ if $i,j \in \mathcal{I}_{F}$ and $0$ otherwise, 
where $\mathcal{I}_{F} = \left\lbrace i \vert \mathbf{M}^k_{x_i,y_i} = 1  \right\rbrace$, $\left(x_{i}, y_{i}\right)$ is the center pixel coordinates of the $i-$th patch in the image, located $\dfrac{P}{2}$ pixels (assuming $P$ even) from the top-left corner of the patch in both vertical and horizontal directions. We define the background correlation values as $\mathbf{\ddot C}^{k} = \mathbf{C}^{k} - \mathbf{\dot C}^{k}$. 
Note that $\mathbf{\dot C}^{k}$ contains correlations between \textit{only} features in the foreground, while $\mathbf{\ddot C}^{k}$ contains correlations between features from both the foreground and the background.

More formally, our proposed self-supervised approach consist of minimizing the distance between a given image $\mathbf{X}^{k}$ and an estimated image obtained by decoding the LGR $\mathcal{P}^k=\left( \mathbf{F}^{k}, \mathbf{W}^{k} \right)$ using a GCN denoted by $\mathcal{A}_{\boldsymbol{\bar{\Omega}}}$ with trainable parameters $\boldsymbol{\bar{\Omega}
}$, and a sparse regularization term differentiating between foreground and background. The proposed joint minimization problem is 

\begin{equation}
    \begin{split}
        \boldsymbol{\Theta}^*, \boldsymbol{\bar{\Omega}^*} = & \argmin_{\boldsymbol{\Theta}, \bar{\Omega}} \dfrac{1}{2} \sum_{k} \Vert \mathbf{x}^{k} - \mathcal{A}_{\bar{\Omega}} \left( \mathbf{F}^{k}, \mathbf{W}^{k} \right) \Vert_{F}^{2} \\
        & + \sum_{k} {\alpha \Vert \mathcal{R}_{\boldsymbol{\Theta}} \left( \mathbf{\dot C}^{k} \right) \Vert_{1} + \beta \Vert \mathcal{R}_{\boldsymbol{\Theta}} \left( \mathbf{\ddot C}^{k} \right) \Vert_{1}},
    \end{split}
    \label{Eq:loss_function1}
\end{equation}
where ${\mathbf{x}^k} = \text{vec} \left( {\mathbf{X}}^{k} \right)$ is the vector form of the ground-truth image,  ${\bar{\Omega}} = \left\lbrace \boldsymbol{\Omega}^{h} \right\rbrace_{h=1}^{H}$ is the set of trainable parameters for $H$ graph convolutional layers (GCLs), and $\alpha$ and $\beta$ are regularization parameters that control the sparsity strength of the selection matrices $\mathcal{R}_{\boldsymbol{\Theta}} \left( \mathbf{\dot C}^{k} \right)$ and $\mathcal{R}_{\boldsymbol{\Theta}} \left( \mathbf{\ddot C}^{k} \right)$. 
We give more importance to the foreground by choosing $\beta>\alpha$.
Mathematically, the output for the $h$-th GCL and the $i$-th vertex of the GCN $\mathcal{A}_{\bar{\Omega}} \left( \cdot \right)$ that build an estimation of the $k$-th image from its graph presentation is
\begin{equation}
    \mathbf{f}^{k(h)}_{i} = \sigma\left(\mathbf{U}^{h} \ \mathbf{f}^{k(h-1)}_{i} + \mathbf{B}^{h} \sum_{j \in \mathcal{J} (i)} \mathbf{W}^{k(h)}_{i,j} \ \mathbf{f}^{k(h-1)}_{j}\right),
    \label{Eq:loss_function2}
\end{equation}
where $\Omega^{h} = \left( \mathbf{U}^{h} \in \mathbb{R}^{D^{h} \times D^{h-1}}, \mathbf{B}^{h} \in \mathbb{R}^{D^{h} \times D^{h-1}} \right)$ is the tuple of learnable matrices for the $h$-th GCL, $\mathcal{J} (i)$ denotes the neighborhood around the $i$-th vertex, $\mathbf{W}^{k(h)}$ is the edge weights for the $h$-th layer, and $\sigma$ denotes a component-wise non-linear function. 
For the first layer ($h=1$) the values $\mathbf{f}^{k(0)}_i$ correspond to the ViT of the $i$-th patch, and $\mathbf{W}^{k(0)}$ is given by \eqref{eq:weight_matrix}. The input and output feature sizes are given by $D^{0} = D$ and $D^{H} = 1$. 

Additionally, since vertices in our LGR are constructed from non-overlapping patches of the image, processing the graph will result in graphs/images of lower resolution ($M/P\times N/P$ nodes/pixels). 
To generate images of the same size as those in the training dataset, we apply multiple stages of $2\times$ upsampling starting with the output of the last $\log_{2}(P)$ GCL, so that, at the final layer, the number of vertices matches the number of pixels in the original images. In each of these stages of $2\times$ upsampling, for each existing vertex we insert three additional vertices positioned to the right, below, and right-below, and compute the unknown feature values using an inverse-distance weighted interpolation from the $K_\text{int}$ known nearest neighboring features.
The interpolated feature for the new $n$-th vertex at position $\mathbf{p}_{n} = \left( x_n, y_n \right)$ in the $h$-th GCL denoted as $\mathbf{f}_{n}^{k(h)}$ is 
\begin{equation}
    \mathbf{f}_{n}^{k(h)} = \sum_{i \in \mathcal{N}(n)}w_{\text{int}} \left( i \right) \mathbf{f}_{i}^{k(h)},
    \label{eq:interpolation}
\end{equation}
where $\mathcal{N} (n)$ denotes the neighborhood containing the $K_\text{int}$ nearest known neighbors around the vertex to be interpolated ($n$-th vertex) and the  interpolation weights are $w_{\text{int}} \left( i \right) = {1}/{\Vert \mathbf{p}_{i} - \mathbf{p}_{n}\Vert_{2}^{2}}$. Once we obtain higher-resolution features, we recompute the edge weights of the $h$-th layer using the same structure as in \eqref{eq:weight_matrix}, i.e., $\mathbf{W}^{k(h)} =  \mathcal{R}_{\boldsymbol{\Theta}} \left( \mathbf{C}^{k(h)} \right) \circ \mathbf{C}^{k(h)}$, where $\mathbf{C}^{k(h)}= \mathcal{N} \left( \mathbf{F}^{k(h)} \mathbf{F}^{k(h)\top} \right)$, $\mathbf{F}^{k(h)} = \left[\mathbf{f}^{k(h)}_1, \cdots, \mathbf{f}^{k(h)}_{N_h} \right]^\top$, and $N_h$ is the number of vertices of the $h$-th layer. 
Finally, the estimated image $\hat{\mathbf{x}}^{k}$ is built from the vertices of the last GCL $\left[\mathbf{f}^{k(H)}_1, \cdots, \mathbf{f}^{k(H)}_{NM} \right]$.

\begin{figure*}
    \centering
    \includegraphics[width=\linewidth]{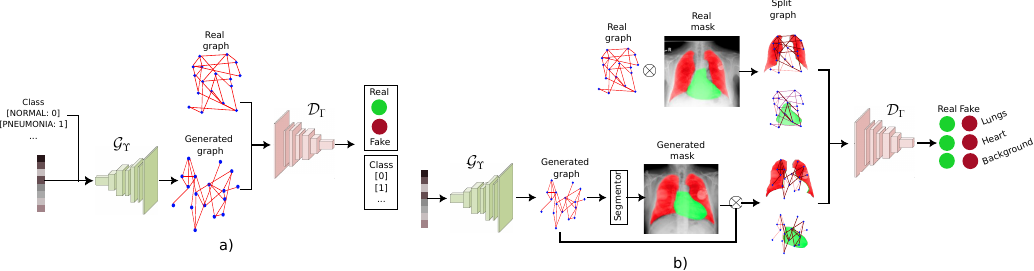}
    \caption{Conditional graph-based GAN architectures for image diagnostic tasks. a) ACGAN architecture: class-conditioned synthesis is learned to generate graphs with class labels, e.g., graphs belonging to the classes NORMAL/PNEUMONIA for the pneumonia classification task. b) SegAN architecture: Domain translation is performed to estimate the segmentation mask from a pre-generated graph where the discrimination process is graph-conditioned.  
    }
    \label{fig:GAN_approaches}
\end{figure*}

\begin{figure*}[t]
	\centering
	\includegraphics[width=\linewidth]{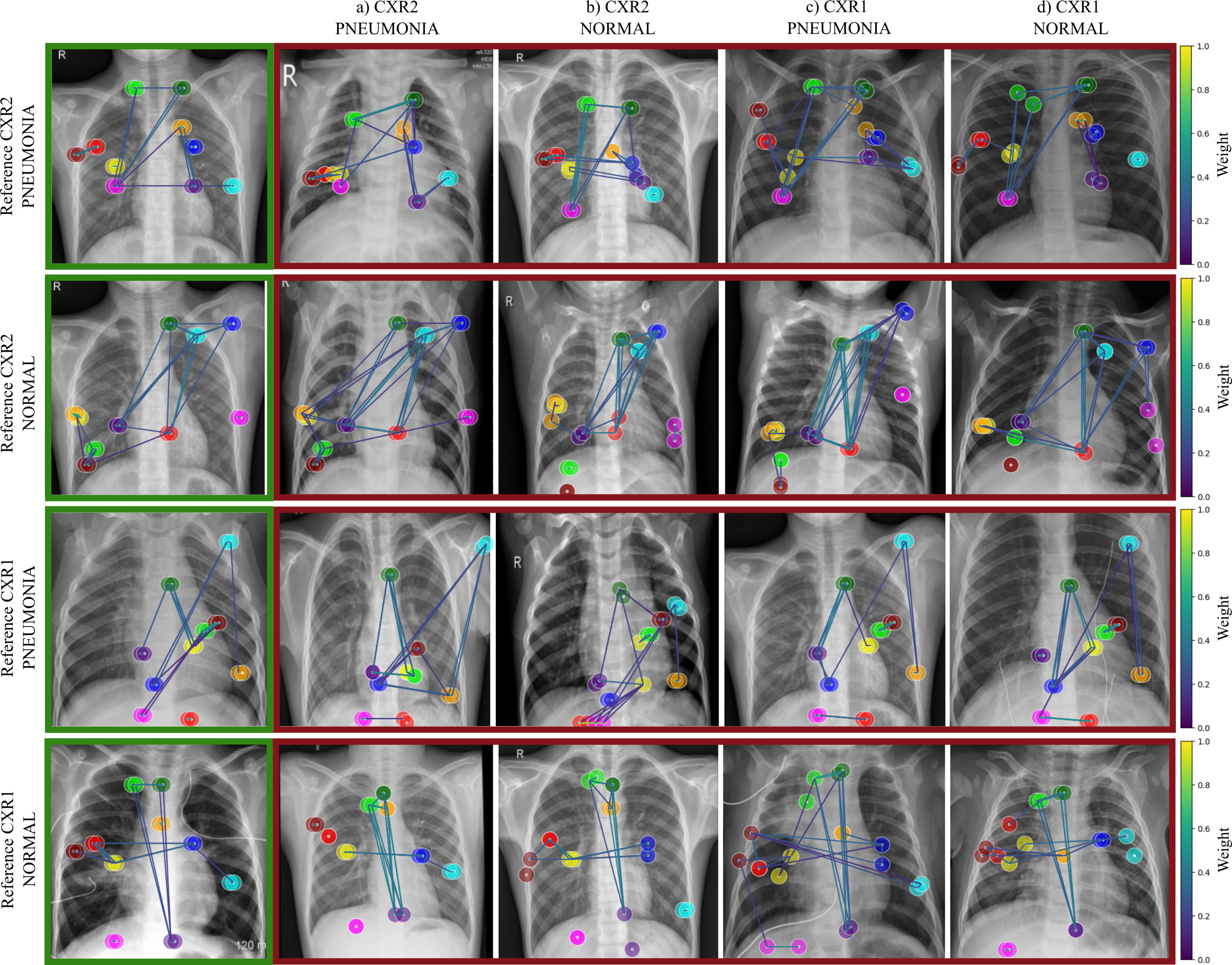}
	\vspace{-10pt}
	\caption{Graph semantic consistency across images.  
In the left-most \textit{reference} image (green box) in each row, chosen from a given dataset and class, we randomly sample ten points and give each a different color. 
To improve the visualization, we added a second point of the same color slightly to the right of each reference image, yielding 20 points per image. 
For the remaining \textit{target} images along a row (red box), each from a different dataset and class, we identify regions having semantic and structural similarity to regions in the reference image in the same row. 
To do so, given the ViT features of a randomly chosen point in the reference image, we find in each target image the point whose ViT features are closest, 
and assign to this target-image point the same color assigned to the matching reference-image point.    
Thus, points of the same color located in different target images along a row indicate similar regions. 
Within each image, sampled points are connected using our learned graph weights. 
These visualizations highlight two key observations: 
(1) regions with similar semantic and structural characteristics across images in the same class tend to preserve a core connectivity structure; and (2) this graph consistency also emerges across images from different datasets and classes (e.g., pneumonia vs. normal), illustrating the robustness of the learned graph semantics.
    }
	\label{fig:consistency}
\end{figure*}


\section{Structure-preserving Adversarial Generation}
We propose guiding data augmentation within the GM framework by leveraging our proposed LGR. To do so, consider the set of LGRs $\mathcal{X}_{\text{LGR}}$ obtained from the training dataset $\mathcal{X}$, i.e., $\mathcal{X}_{\text{LGR}} = \left\lbrace\mathcal{P}^{k}=\left( \mathbf{F}^{k}, \mathbf{W}^{k} \right) \right\rbrace_{k=1}^{K}$. 
We propose an adversarial training of a GM designed to generate a semantic LGR that closely adheres to the LGR distribution observed in $\mathcal{X}_{\text{LGR}}$. 
We approach this problem by learning the signal on the graph distribution $p_{data}(\mathbf{F})$ and constructing LGRs $\mathcal{P} = ~\left( \mathbf{F}, \mathbf{W} \right)$ from samples $\mathbf{F}$ of this distribution and using the CNN $\mathcal{R}_{\boldsymbol{\Theta}^*}$ learned using  \eqref{Eq:loss_function1} along \eqref{eq:weight_matrix} to compute the edge weights $\mathbf{W}$. 
The generator $\mathcal{G}_{\Upsilon}$ and discriminator $\mathcal{D}_{\Gamma}$ networks are optimized via the following \textit{minmax} problem

\begin{equation}
    \begin{split}
        \min_{\Upsilon} \max_{\Gamma} \hspace{2mm} & \mathbb{E}_{\mathbf{F} \sim p_{data} \left( \mathbf{F} \right)} \log{ \left( \mathcal{D}_{\Gamma} \left( \mathbf{F}, \mathbf{W} \right) \right)} + \\ 
        &\mathbb{E}_{\mathbf{z} 
 \sim p \left( \mathbf{z} \right)} \log{\left( 1 - \mathcal{D}_{\Gamma} \left( \mathcal{G}_{\Upsilon} \left( \mathbf{z} \right), \hat{\mathbf{W}} \right) \right)},
    \end{split}
    \label{Eq:loss_function_GAN}
\end{equation}
where $p(\mathbf{z})$ is an unstructured prior distribution in the vector $\mathbf{z} \in \mathbb{R}^b$, $\hat{\mathbf{F}} = \mathcal{G}_{\Upsilon} \left( \mathbf{z} \right)$ denotes the generated signal, $\hat{\mathbf{W}}$ is the weights matrix obtained from $\hat{\mathbf{F}}$ using \eqref{eq:weight_matrix}, and the pair $\Upsilon$ and $\Gamma$ are trainable parameters for the generator and discriminator networks, respectively. 
$\mathcal{D}_{\Gamma}(\mathbf{F},\mathbf{W})$ outputs
a single scalar that represents the probability that the LGR $\mathcal{P} = (\mathbf{F}, \mathbf{W})$ came from training data rather than the generator distribution.  
The generator network $\mathcal{G}_{\Upsilon}(\mathbf{z})$ implicitly defines the probability distribution of the LGR samples obtained when $z\sim p(\mathbf{z})$. Thus, after training, new realizations of random noise $\left\lbrace\mathbf{z}^{q} \right\rbrace_{q=1}^{Q}$ enable the generation of $Q$ news graphs $\mathcal{P}^{q} = \left( \mathbf{F}^{q}, \mathbf{W}^{q} \right)$ and by taking advantage of the pre-trained network $\mathcal{A}_{\bar{\Omega}^*}$ obtained from \eqref{Eq:loss_function1} we generate structure-preserving medical images as

\begin{equation}
    \mathbf{X}^{q} = \mathcal{A}_{\bar{\Omega}^*} \left(\mathbf{F}^{q}, \mathbf{W}^{q} \right).
    \label{Eq:new_Mimages}
\end{equation}
We highlight that $\mathcal{A}_{\bar{\Omega}^*} \left( \cdot \right)$ jointly optimized with $\mathcal{R}_{\boldsymbol{\Theta}^*} \left( \cdot \right)$ works as a self-supervised decoder that captures how graphs relate to images since it was trained as part of an encoder-decoder pipeline (image-to-graph and graph-to-image). 

\subsection{Using graph-based GMs for high-level tasks}

We also exploit the rich structure of our proposed graph-based GM to perform data augmentation
for training graph-based learning models for \textit{classification} and \textit{segmentation} of chest X-ray images. 
For data augmentation, we need to generate not only reliable graphs but also their associated class labels and segmentation maps, which we generate using conditional GAN models tailored to each task. 

For \textit{classification}, we employ an auxiliary classifier GAN architecture (ACGAN) to obtain both the chest X-ray graph $\mathcal{P}^{q}$ and its pneumonia classification label $t^{q}$. In the ACGAN model illustrated in \autoref{fig:GAN_approaches}(a), the input to the generator $\mathcal{G}_{\Upsilon}$ is the concatenation of $\mathbf{z}$ and a corresponding class label $t$ while the outputs of the discriminator are the probability that the input LGR came from the real dataset and the probability distribution over class labels \cite{odena2017conditional}. 
The generated augmentation samples and the original classification dataset are combined in an augmented training dataset $\mathcal{D}_{\text{aug}}=\left\lbrace \left( \mathcal{P}^{r}, t^{r} \right) \right\rbrace_{r=1}^{K+Q} = \left\lbrace \left\lbrace \left( \mathcal{P}^{k}, t^{k} \right) \right\rbrace_{k=1}^{K} \bigcup \left\lbrace \left( \mathcal{P}^{q}, t^{q} \right) \right\rbrace_{q=1}^{Q} \right\rbrace$ and then used to train a GCN which aims to minimize the following objective function associated with the \textit{classification} problem 

\begin{equation}
    \begin{split}
        \bar{\Delta}^*, \bar{\Phi}^* = \argmin_{\bar{\Delta}, \bar{\Phi}} & - \dfrac{1}{K+Q} \sum_{r=1}^{K+Q} t^{r} \text{log} \left( \mathcal{F}_{\bar{\Delta}} \left( \mathcal{C}_{\bar{\Phi}} \left( \mathcal{P}^{r} \right) \right) \right) \\
        & + \left( 1 - t^{r} \right) \text{log} \left( 1 - \mathcal{F}_{\bar{\Delta}} \left( \mathcal{C}_{\bar{\Phi}} \left( \mathcal{P}^{r} \right) \right) \right),
    \end{split}
    \label{Eq:Classification_problem}
\end{equation}
where $p^{r} = \mathcal{F}_{\bar{\Delta}} \left( \mathcal{C}_{\bar{\Phi}} \left( \mathcal{P}^{r} \right) \right)$ is the predicted classification probability for the $r$-th sample, $\mathcal{C}_{\bar{\Phi}} \left( \cdot \right)$ is the GCN-based classification network with trainable parameters $\bar{\Phi}$ and $\mathcal{F}_{\bar{\Delta}} \left( \cdot \right)$ is a fully-connected network with softmax output probabilities and trainable parameters $\bar{\Delta}$. 

For the \textit{segmentation} task, based on the segmentation adversarial neural network architecture (SegAN) \cite{xue2018segan}, we propose a graph-conditioned GAN,  referred to as graph-SegAN,  to obtain both the chest X-ray graph $\mathcal{P}^{q}$ and its segmentation map $\mathbf{M}^{q} \in \mathbb{Z}^{N \times M}$. In this architecture, illustrated in \autoref{fig:GAN_approaches}(b), the generator, conditioned on pre-generated LGRs obtained using an unconditional GAN, generates LGRs and probability label maps. The discriminator is designed to extract hierarchical features from the segmented image and evaluate the realism of the LGR.
Combining  generated and original dataset we obtain the augmented dataset $\mathcal{D}_{\text{aug}}=\left\lbrace \left( \mathcal{P}^{r}, \mathbf{M}^{r} \right) \right\rbrace_{r=1}^{K + Q}$ to train our DL model for the segmentation task. The segmentation maps are represented in categorical format as $\mathbf{M}^{r} = \sum_{l=1}^{L} l \cdot \bar{\mathbf{M}}^{r(l)}$ where $\bar{\mathbf{M}}^{r(l)} \in \left[ 0, 1 \right]^{N \times M}$ for all class labels $l \in \left[ 1, 2, ... L \right]$. 
A \textit{segmentation} model, based on a GCN $\mathcal{S}_{\bar{\Lambda}} \left( \cdot \right)$
with trainable parameters $\bar{\Lambda}$ is trained to minimize the multi-class cross-entropy loss
\begin{equation}
    \begin{split}
        \bar{\Lambda}^* = \argmin_{\bar{\Lambda}} & - \dfrac{1}{T} \sum_{r=1}^{K + Q} \sum_{n, m=1}^{N, M} \sum_{l=1}^{L} \bar{\mathbf{M}}^{r(l)}(n,m) \ \text{log} \left( \mathcal{S}_{\bar{\Lambda}} \left( \mathcal{P}^{r} \right) \right)
    \end{split}
    \label{Eq:Segmentation_problem}
\end{equation}
where dividing by $T = \left( K + Q \right)MN$, the number of instances, computes the average cost, and  $\mathcal{S}_{\bar{\Lambda}} \left( \mathcal{P}^{r} \right) = \hat{\mathbf{M}}^{r(l)}(n, m)$ is the predicted segmentation probability for the $r$-th sample, $l$-th channel and pixel position $(n,m)$. 

\section{Datasets and Implementation} 

\subsection{Datasets}

\begin{table}[t]
\centering
\caption{Summary of Datasets}
\resizebox{\columnwidth}{!}{%
\begin{tabular}{|l|c|c|}
\hline
\textbf{Dataset} & \textbf{Train/test} & \textbf{Class Distribution} \\
\hline
CXR1 & 5,232/624 & 4,273 Pneumonia, 1,583 Normal \\
\hline
CXR2 & 5,550/1,389 & 2,313 each (COVID-19, Pneumonia, Normal) \\
\hline
JSRT & 199/48 & Lungs, Heart, Background \\
\hline
\end{tabular}}
\label{tab:datasets}
\end{table}

For the classification assessment, we employ two chest X-ray image datasets: 
1) The ``CXR1'' dataset from the Guangzhou Women and Children's Medical Center in China (pneumonia detection) \cite{kermany2018identifying}, and  
2) The \text{COVID19 Pneumonia Normal Chest Xray PA Dataset} (``CXR2", pneumonia and COVID-19 detection)\footnote{https://www.kaggle.com/datasets/amanullahasraf/covid19-pneumonia-normal-chest-xray-pa-dataset}.  
For the lung segmentation assessment, we employ the Japanese Society of Radiological Technology (JSRT) dataset with ground truth lung segmentations \cite{shiraishi2000development}, which has been widely used for tasks such as lung nodule detection and lung segmentation. 
Additional information about these datasets can be found in \autoref{tab:datasets}.
Images from ``CXR1", ``CXR2", and ``JSRT" were used in the training dataset 
for (i) learning the LGR (i.e., learning parameters ${\boldsymbol{\Theta}}$) and (ii) the GCN for reconstruction  (i.e., learning parameters ${\boldsymbol{\bar{\Omega}}}$) based on the self-supervised joint minimization problem in \eqref{Eq:loss_function1}. 
Once we have learned $\mathcal{R}_{\boldsymbol{\Theta}}$, we can build the \textit{training} set of LGRs $\mathcal{X}_{\text{LGR}}$ of images in ``CXR1", ``CXR2", and ``JSRT" by using their ViT features and \eqref{eq:weight_matrix}. Then, depending on the experiments in the next section, we can use these LGRs for training graph-based GANs or GCNs. In the case of conditional graph-based GANs, each LGR is paired with the label of the image it was generated from and used as input to the graph-based model. 

\subsection{Metrics} For quantitative evaluation, we use the Fréchet inception distance (FID) to evaluate the X-ray image synthesis \cite{yu2021frechet}. Specifically, the FID metric is employed to measure the distance between the distributions of real images $\mathcal{X} = \left\lbrace \mathbf{X}^{k} \right\rbrace_{k=1}^{K}$ and generated images $\hat{\mathcal{X}} = \left\lbrace \hat{\mathbf{X}}^{q} \right\rbrace_{q=1}^{Q}$. Inception embeddings are assumed to be two multivariate normal distributions. FID is defined as
\begin{equation}
    \text{FID}(\mathbf{X},\hat{\mathbf{X}}) = \Vert \mu_{x} - \mu_{\hat{x}} \Vert^{2} - TR \left( \Sigma_{x} + \Sigma_{\hat{x}} - 2 \sqrt{\Sigma_{x} \Sigma_{\hat{x}}} \right)
    \label{Eq:FID}
\end{equation}
where $\left(\mu_{x}, \mu_{\hat{x}} \right)$ and $\left(\Sigma_{x},\Sigma_{\hat{x}}\right)$ are tuples representing the magnitudes and covariance of the embeddings \cite{yu2021frechet}. 
We use the accuracy (ACC) metric, area under the ROC curve (AUC), and F1 score (F1) for evaluating the image classification task (no pneumonia/pneumonia). As in SOTA, the DICE metric is used to evaluate the segmentation task. 
Taking the probability for the pixel $\left( n, m \right)$ in the $l$-th class $\mathbf{P}^{k(l)}(n, m)$ as the network output on the channel $l$ before the softmax activation function, the DICE metric can be calculated as 
\begin{equation}
    \text{DICE} \left( \mathbf{X}^{k(l)}, \mathbf{M}^{k(l)} \right) = 2 \dfrac{\vert \bar{P} \left(\mathbf{X}^{k(l)}\right) \cap \bar{R} \left( \mathbf{M}^{k(l)} \right) \vert}{\vert \bar{P} \left( \mathbf{X}^{k(l)} \right) + \bar{R} \left(\mathbf{\mathbf{M}}^{k(l)} \right) \vert}
    \label{Eq:DICE}
\end{equation}
where $\bar{R} \left( \mathbf{M}^{k(l)} \right)= \left\lbrace \left( n, m \right): \mathbf{M}^{k(l)}(n, m) = 1 \right\rbrace$ and $\bar{P} \left( \mathbf{X}^{k(l)} \right)= \left\lbrace \left( n,m \right): \vert \mathbf{P}^{k(l)}(n, m) - 1 \vert < \epsilon \right\rbrace$  
are the sets of pixels outside of the background that belong to the $l$-th class for the predicted and real segmentation masks, respectively.

\section{Results} 
We conduct several ablation studies and comparisons to evaluate the contributions of our LGR for X-ray data augmentation in high-level tasks, such as classification and lung segmentation using GCNs.
First, we analyze the LGR in terms of the best construction strategy of the graph topology $\left( \mathbf{W} \right)$, the best GAN architecture for LGRs generation, and GCN architectures for image synthesis, classification, and lung segmentation. 
Second, we compare the quality of our proposed generated X-ray images as augmentations with traditional augmentations and SOTA GANs by feeding them into classification and lung segmentation baseline networks. Finally, we compare our learned LGR against other SOTA LGR representations for the classification and lung segmentation tasks using GCNs.

For all the experiments, some parameters were fixed as follows: patch size $P = 16$, image size $M, N = 256$, GCLs $H=10$, and number of neighbors in the graph interpolation $K_{\text{int}}=12$. All our models are implemented in Pytorch and trained on an NVIDIA GeForce RTX 3090 GPU with 24 GB of memory using the ADAM optimizer. 

\subsection{Analysis of the LGR construction for image generation} 
\begin{table}[ht]
\centering
\caption{Summary of learning strategies for constructing the weights matrix $\mathbf{W}$.}
\begin{tabular}{c |l | l}
\toprule
\textbf{Strategy} & \textbf{Weight Matrix Definition} & \textbf{Sparsity Regularization} \\
\midrule
\textit{Setup 1} & $\mathbf{W} = \mathcal{R}_{\boldsymbol{\Theta}}(\mathbf{C})$ & $\alpha = \beta = 0 $\\
\textit{Setup 2} & $\mathbf{W} = \mathcal{R}_{\boldsymbol{\Theta}}(\mathbf{C}) \cdot \mathbf{C}$ & $\alpha = \beta = 0$ \\
\textit{Setup 3} & $\mathbf{W} = \mathcal{R}_{\boldsymbol{\Theta}}(\mathbf{C}) \cdot \mathbf{C}$ & $\alpha = \beta \neq 0$ \\
\textit{Setup 4} & $\mathbf{W} = \mathcal{R}_{\boldsymbol{\Theta}}(\mathbf{C}) \cdot \mathbf{C}$ & with foreground prioritized \\
\bottomrule
\end{tabular}
\label{tab:learning_strategies}
\end{table}

\noindent \textbf{Finding the latent graph}.
To examine our proposed latent graph construction, we consider four different strategies to learn $\mathcal{R}_{\boldsymbol{\Theta}}$, and hence the matrix $\mathbf{W}$ (see \autoref{tab:learning_strategies}). In the first two experiments, we do not use sparsity regularization
Specifically, in the first strategy, we just use binary weights that connect ($1$) or disconnect ($0$) features on the graph, such that $\mathbf{W} = \mathcal{R}_{\boldsymbol{\Theta}}(\mathbf{C})$. 
Second, we use real weights obtained as the product of correlation values $\mathbf{C}$ and the binary selection matrix $\mathcal{R}_{\boldsymbol{\Theta}}(\mathbf{C})$, 
such that $\mathbf{W} = \mathcal{R}_{\boldsymbol{\Theta}}(\mathbf{C}) \cdot \mathbf{C}$. In the last two learning strategies, we promote sparsity of the weight matrix $\mathbf{W} = \mathcal{R}_{\boldsymbol{\Theta}}(\mathbf{C}) \cdot \mathbf{C}$. 
Thus, in the third strategy, we minimize \eqref{Eq:loss_function1} giving the same importance to background and foreground by setting $\alpha = \beta$. Finally, in the last learning strategy, we prioritize the connection between the features inside the lung and heart areas segmented as shown in  \autoref{fig:segmentation}. 
To this end, we solve the proposed optimization problem in \eqref{Eq:loss_function1} with regularization parameters such that $\beta=12\alpha$. 

Each setup is evaluated for both reconstruction quality in terms of PSNR in dB and ``No. edges,'' denoting the number of edges preserved on the graph. All experiments use the baseline GCN GraphSAGE for reconstruction. The reported values correspond to the average of the CXR1, CXR2, and JSRT test datasets. Besides, we include the result considering \textit{only} the foreground region. The quantitative results are reported in \autoref{tab:ablation_graph}, showing that the sparsity-promoting approaches  $\mathcal{R}_{\boldsymbol{\Theta}}(\mathbf{C}^k) \cdot \mathbf{C}^{k}$ achieve better PSNR performance than the other learning strategies using a similar number of edges in the graph. 
Further, separately promoting sparsity in foreground and background enables better representation for the lung and heart areas. According to the results, the sparsity-promoting approaches are adopted for the remaining results section, including the evaluation of the latent graph for the classification and lung segmentation tasks. 
\begin{table}[t]
    \centering
    \caption{Graph representation results in terms of reconstruction quality (PSNR in \textnormal{dB}) and number of edges for learning strategies of $\mathbf{W}$ described in \autoref{tab:learning_strategies}. Columns labeled (\textnormal{Foreg}) report values computed using only the foreground.}
    \resizebox{\columnwidth}{!}{%
    \begin{tabular}{c|c|c|c|c}
        \toprule
        \textbf{Setup} & \textbf{PSNR } & \textbf{PSNR (Foreg)} & \textbf{No. Edges } & \textbf{No. Edges (Foreg)} \\
        \midrule
        \textit{Setup 1} & 27.05 & 26.81 & \textbf{353556} (7.73\%) & 63522 (1.39\%) \\
        \textit{Setup 2} & 30.01 & 29.12 & 347197 (7.59\%) & 62841 (1.34\%) \\
        \textit{Setup 3} & 32.98 & 30.57 & 347797 (7.60\%) & 89395 (1.95\%) \\
        \textit{Setup 4} & \textbf{34.06} & \textbf{35.05} & 352411 (7.70\%) & \textbf{205600} (4.50\%) \\
        \bottomrule
    \end{tabular}}
    \label{tab:ablation_graph}
\end{table}

\begin{figure}[t]
	\centering
\includegraphics[width=\linewidth]{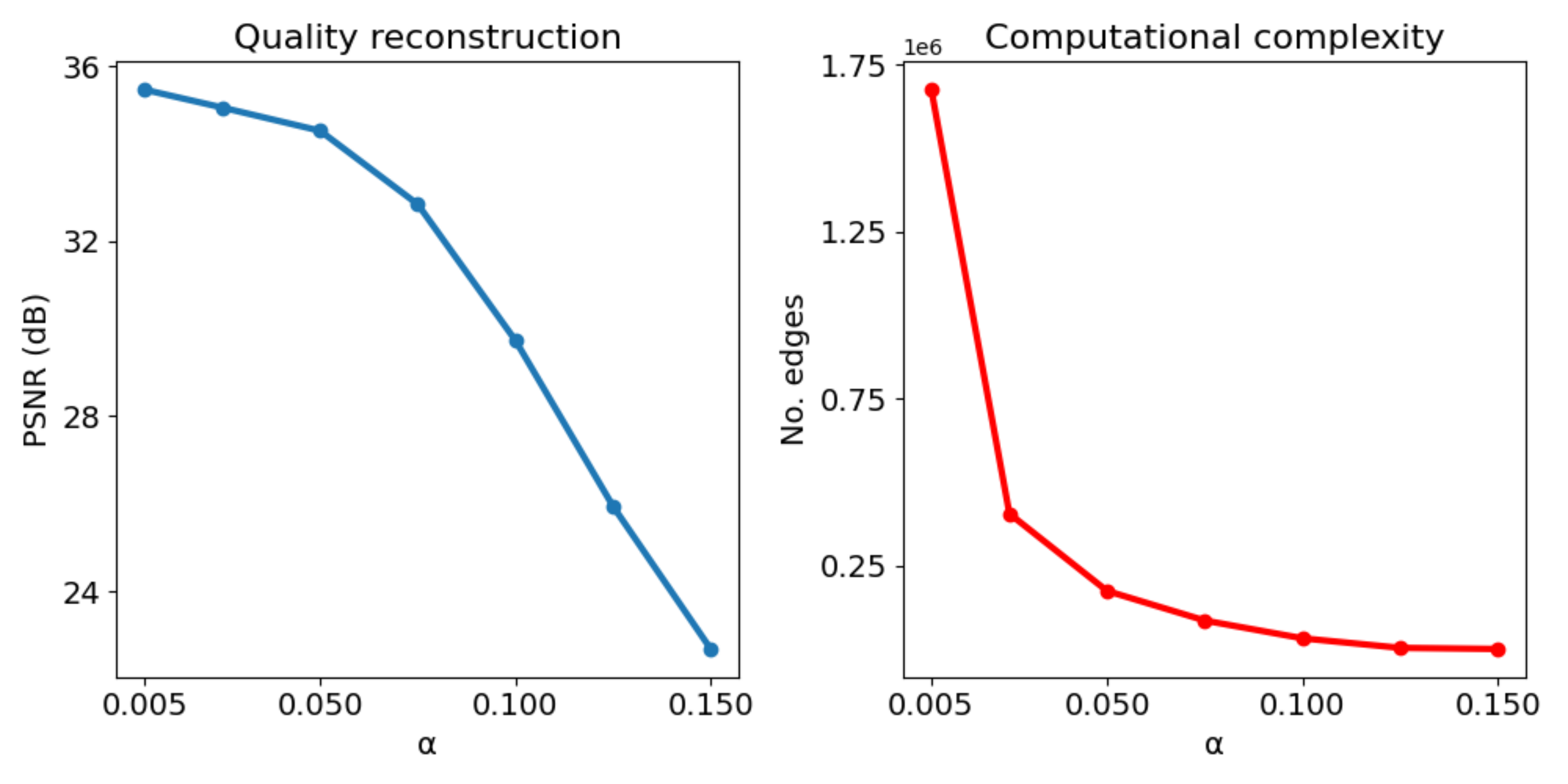}
	\vspace{-10pt}
	\caption{Image quality in terms of PSNR and computational complexity indicating the average number of connected edges inside graphs for different values of the sparsity parameter $\alpha$.}
	\label{fig:sparsity}
\end{figure}

\noindent \textbf{Sparse graph analysis}. Here, we further evaluate the role of sparsity in the proposed approach in terms of computational complexity and reconstruction quality. By promoting sparsity on the selection matrix and setting a high percentage of edge coefficients to zero, we alleviate the computational complexity of training when compared to processing the full-connected graph. However, there is a trade-off between graph computational complexity and reconstruction quality, e.g., sparsely connected graphs decrease the image quality, and graphs with a greater number of edges increase both the reconstruction quality at the cost of higher computational complexity. We evaluate in \autoref{fig:sparsity} the effect of varying the sparsity regularization parameters $\alpha$ and $\beta$ on the quality of reconstruction and computational complexity. For simplicity in the analysis, we choose $\beta = 12 \alpha$. The average PSNR and number of edges for the values $\alpha = 0.005, 0.025, 0.05, 0.075, 0.1, 0.125, 0.15$ are shown on the left and right of \autoref{fig:sparsity}, respectively. As shown in this figure, the graphs with a very high quality reconstruction may require an impractical number of edges for processing through GCNs, e.g., reconstruction qualities close to $36$ dB require around 1.8 millions of edges. For our experiments, we find that a good trade-off is achieved with $\alpha=0.025$ with an average PSNR over $35$ dB and a feasible number of edges $\ll 1e^{+6}$ for processing with GCNs.  
Therefore, the remaining experiments for graph latent construction below are carried out under the setup $\alpha = 0.025$. 

\begin{table}[t]
	\centering
	\caption{Image reconstruction results in terms of PSNR (\textnormal{dB}) for different GCNs receiving as input the learned LGR from \textnormal{\textit{Setup 3}} and \textnormal{\textit{Setup 4}}.}
\resizebox{\columnwidth}{!}{%
	    \begin{tabular}{c|c|c|c|c|c}
		\toprule
		\multirow{2}{*}{}{\small $\mathbf{W}$}  &
		\multirow{2}{*}{}{\small GraphSAGE} &
		\multirow{2}{*}{}{\small GIN} &
		\multirow{2}{*}{}{\small SAGPool} &
		\multirow{2}{*}{}{\small EdgePool} &
        \multirow{2}{*}{}{\small GAT}
		\\ \hline
        \textit{Setup 3} 
		 & 32.98 & 34.55  & 35.01 & 35.11 & \textbf{35.94}
		\\ 
		\textit{Setup 4 }               
		 & 34.06 & 34.58 & 35.12 & 35.09 & \textbf{36.14} \\ \bottomrule
\end{tabular}}
\label{tab:reconstructionGCN}
\end{table}

\begin{figure*}[t]
	\centering
	\includegraphics[width=0.95\linewidth]{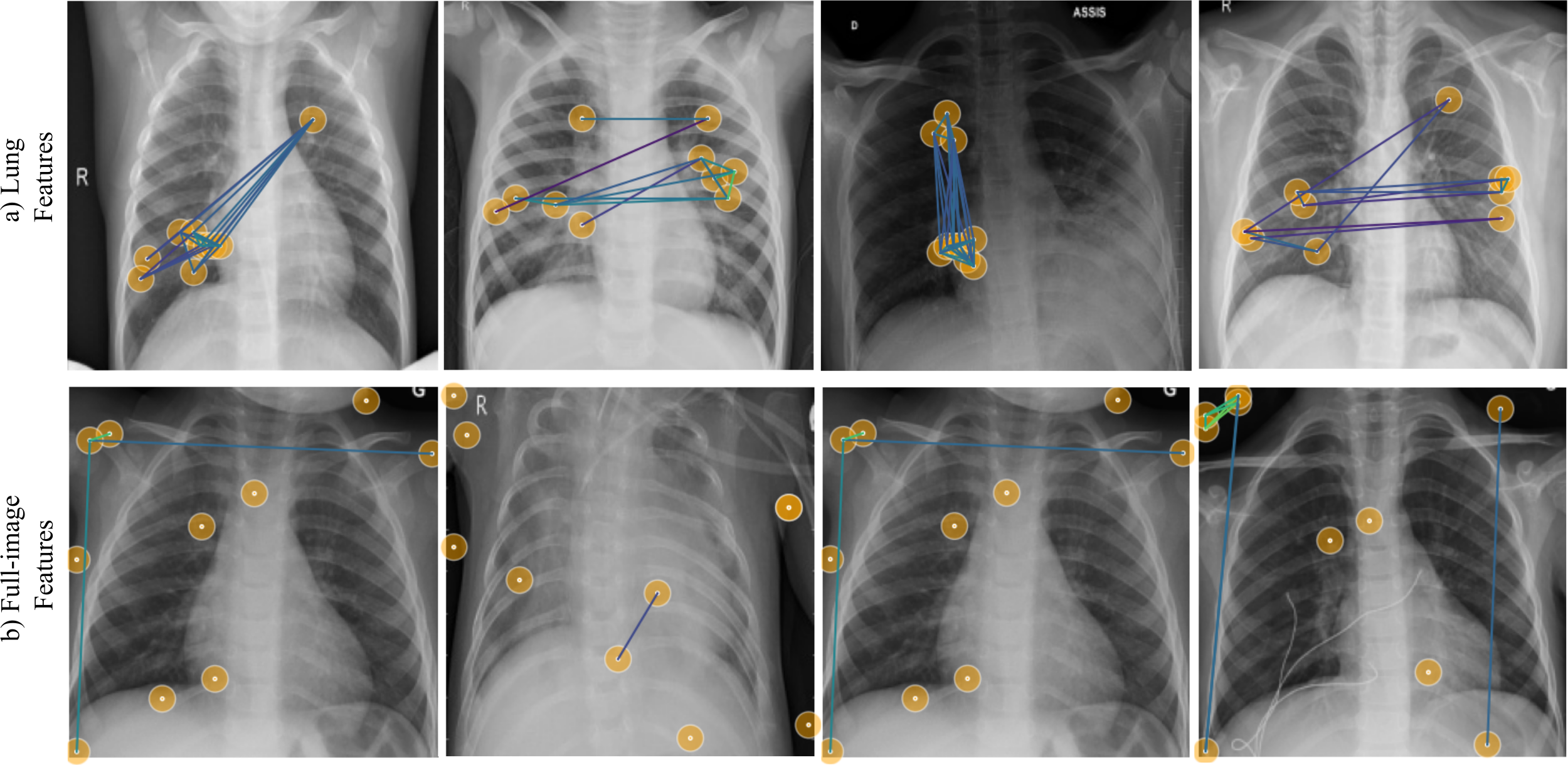}
	\vspace{-10pt}
	\caption{Graph concordance: For different X-ray images, we sample ten different points in random locations within the lungs (top row) and manually select points in diverse anatomical regions (bottom row). Within each image, sampled points are colored differently and connected using our learned graph weights. These visualizations show denser connections between features belonging to similar anatomical regions, most notably within the lungs (top row). In contrast, sparser connections and even near-complete disconnection occur between dissimilar regions (bottom row), highlighting the anatomical concordance of the learned feature relationships, where connections are denser and more coherent among features from similar spatial regions.} 
\label{fig:concordance}
\end{figure*}

\noindent \textbf{Semantic
and structural relationships}. 
We evaluate our LGR construction strategy on how well it preserves the structural information of medical images by 1) verifying whether the graph captures semantic relationships that are consistent throughout the entire dataset, and 2) its ability to connect anatomically similar regions, while disconnecting anatomically distinct regions. 
Thus, we identify regions corresponding to the same anatomical part across different images from feature correspondence as in \cite{amir2021deep}. We randomly sample points and plot each of them with a different color for \textit{reference} images belonging to the classes CXR1 PNEUMONIA, CXR1 NORMAL, CXR2 PNEUMONIA, and CXR2 NORMAL as shown on the first green column in \autoref{fig:consistency}. Then, for the remaining \textit{target} images along each red row corresponding to the classes CXR2 PNEUMONIA \autoref{fig:consistency}(a), CXR2 NORMAL \autoref{fig:consistency}(b), CXR1 \autoref{fig:consistency}(c), and CXR1 NORMAL \autoref{fig:consistency}(d), we identify regions having semantic and structural similarity.
To this end, given the ViT features of a randomly
chosen point in the reference image, we find in each target image the point whose ViT features are closest, and assign the same color to the matching reference-image point.  
Thus, the points with the same color along a row indicate similar features. \autoref{fig:consistency} also shows the connection strength between the selected features corresponding to the learned graph weights. 
These visualizations highlight two key observations, (1) regions with similar semantic and structural characteristics across images in the same class tend to preserve a core connectivity structure; and (2) this graph consistency also emerges across images from different datasets and classes (e.g., pneumonia vs. normal), illustrating the robustness of the learned graph semantics.
This establishes that the learned graph captures relationships with a semantic sense, e.g., connections between points that represent the same anatomical regions tend to be preserved throughout most of the images. 
These relationships, while not universally consistent, reflect a level of anatomical correspondence that supports the ability of the constructed graph. 
Finally, we analyze the learned weights in \autoref{fig:concordance} to evaluate the graph concordance, e.g., to connect similar and disconnect distinct anatomical regions. For this, we randomly sample points inside the lungs \autoref{fig:concordance}(a) and manually select points of different anatomical regions \autoref{fig:concordance}(b). Consistently, the graph densely connects points inside the same anatomical region, such as the lungs. In contrast, features of different anatomical regions are mostly disconnected, confirming that the learned graph also captures their dissimilarity.

\noindent\textbf{GCN architectures for reconstruction} To synthesize structure-preserving images, we use GCN architectures based on spatial graph analysis to leverage the semantic information captured by the learned topology. We compare in \autoref{tab:reconstructionGCN} the performance of (GraphSAGE) \cite{hamilton2017inductive}, graph isomorphism network (GIN) \cite{xu2018powerful}, self-attention graph pooling (SAGPool) \cite{lee2019self}, (EdgePool) \cite{diehl2019towards}, and (GAT) \cite{velickovic2017graph} to reconstruct the image from the learned LGR resulting from \textit{Setup 3} and \textit{Setup 4}. These GCN architectures have been properly adapted by adding layers at the end of the network to match the spatial dimensions of the output image required for the reconstruction task.
The GAT network shows the best performance in mapping the latent graph to the corresponding image in both setups. In the remaining experiments, we use the learned LGR using \textit{Setup 4} and the GAT network for generating our graph-based augmentations. 

\noindent \textbf{GAN architecture} We addressed data augmentation using a GM that follows the learned latent graph distribution of the training dataset $\mathcal{X}_{\textnormal{LGR}}$.  
We now compare different GMs that generate LGRs for training GCN architectures employed in classification and segmentation tasks. Thus, we evaluate the generation quality by classification accuracy and DICE metric obtained by training the respective GCN using the augmented dataset from the corresponding GM.
We train these GMs using LGRs from images in ``CXR2"  and ``JSRT", along with their labels, for classification and segmentation tasks, respectively. We use GIN, SAGPool, and gated graph sequence (GGS) neural network for classification.
For segmentation, we use GraphSAGE and GAT for segmentation. 
We train these GCNs using a corresponding augmented dataset using the learned GM for each task with a different number of augmentations $Q=0$, $100$, $300$, $500$, and $1000$. The value $Q=N$ corresponds to the learning with the LGRs of the original dataset plus $N$ augmentations from the GM. The value $Q=0$ corresponds to the learning with the LGRs of the original dataset without any augmentations from the GM. For the  \textit{classification} task, we use DCGAN and WGAN architectures as GM to independently generate LGRs from images belonging to the normal, pneumonia, and COVID-19 classes. 
Besides, we used the ACGAN for the conditional generation of LGRs from the three classes: normal, pneumonia, and COVID-19. For \textit{segmentation}, we also use WGAN and the conditional SegAN to generate the LGR and segmentation mask. From  \autoref{fig:ablationGCNs}, we see that the best results are consistently obtained using the conditional generation approaches ACGAN and SegAN for the classification and segmentation tasks, respectively. For all values of $Q$, SagPool obtains the best classification accuracy, while GAT obtains the best DICE values in segmenting the X-ray images.  

\begin{table*}[t]
\centering
\caption{Classification (ACC) and segmentation (DICE) results for SOTA GCNs trained with the original training set of LGRs plus $Q$ of our generated LGRs using different generative networks. 
$Q=0$ corresponds to training with \textit{only} the LGRs of the training dataset, i.e., training without augmentations. 
}
\scriptsize
\begin{tabular*}{\textwidth}{|@{\extracolsep{\fill}}|p{1.30cm}|p{1.22cm}|p{0.67cm}|p{0.71cm}p{0.71cm}p{0.71cm}|p{0.71cm}p{0.70cm}p{0.71cm}|p{0.71cm}p{0.70cm}p{0.71cm}|p{0.71cm}p{0.70cm}p{0.71cm}|}
    \hline
     \centering {\textbf{Task}} & \centering {\textbf{GCN net}} & \centering {$Q$=0} & \multicolumn{3}{c|}{$Q$=100} & \multicolumn{3}{c|}{$Q$=300} & \multicolumn{3}{c|}{$Q$=500} & \multicolumn{3}{c|}{$Q$=1000} \\
     \hline
    \multicolumn{3}{|c|}{}  & \centering DCGAN & \centering WGAN & \centering \centering ACGAN & \centering DCGAN & \centering WGAN & \centering ACGAN & \centering DCGAN & \centering WGAN & \centering ACGAN & \centering DCGAN & \centering WGAN & ACGAN                \\ \hline

\multirow{3}{*}{\textit{Classification}} & \centering GIN & \centering 95.87& \centering 95.94 & \centering 96.06 & \centering 96.11 & \centering 96.01 & \centering 96.19 & \centering 96.20 & \centering 96.07  & \centering 96.45 & \centering 96.56 & \centering 96.11 & \centering 96.52 & 96.65                     \\  
\multicolumn{1}{|c|}{}                                & \centering SAGPool  & \centering 97.52 & \centering 97.64 & \centering 97.70 & \centering \textbf{98.17} & \centering 97.77 & \centering 98.00 & \centering \textbf{98.34} & \centering 97.90 & \centering 98.30 & \centering \textbf{98.58} & \centering 97.96 & \centering 98.41  & \textbf{98.66}                    \\  
\multicolumn{1}{|c|}{}                                & \centering GGS  & \centering 94.50 & \centering 94.67 & \centering 94.72 & \centering 94.74 & \centering 94.74 & \centering 94.88 & \centering 94.96 & \centering 94.82 & \centering 95.01 & \centering 95.07 & \centering 94.84 & \centering 95.09 & 95.11                    \\ \hline
\end{tabular*}

\begin{tabular*}{\textwidth}{|@{\extracolsep{\fill}}|p{1.30cm}|p{1.20cm}|p{0.65cm}|p{1.25cm}p{1.25cm}|p{1.25cm}p{1.25cm}|p{1.25cm}p{1.25cm}|p{1.25cm}p{1.25cm}|}
         \multicolumn{3}{|c|}{}    & \centering WGAN & \centering SegAN & \centering WGAN & \centering SegAN & \centering WGAN & \centering SegAN & \centering WGAN &  \parbox[c]{1.4cm}{\centering SegAN} \\
        \hline
        \multirow{2}{*}{\textit{Segmentation}} & \centering GraphSAGE & \centering 97.95 & \centering 98.36 & \centering 98.38 & \centering 98.62 & \centering 98.79 & \centering 98.80 & \centering 98.97 & \centering 98.83 & \parbox[c]{1.4cm}{\centering 99.04} \\ 
         & \centering GAT & \centering 98.62 & \centering 98.79 & \centering \textbf{98.81} & \centering 98.92 & \centering \textbf{99.12} & \centering 98.98 & \centering \textbf{99.39} & \centering 99.02 & \parbox[c]{1.4cm}{\centering \textbf{99.46}} \\
        \hline
    \end{tabular*}
    \label{fig:ablationGCNs}
\end{table*}

\subsection{Image Diagnostics Results from Synthetic Images}
In the previous subsection, we evaluated data augmentation in terms of the quality of generated LGRs for GCNs. Here, we address data augmentation in terms of the synthesized images using our generated LGRs and the reconstruction network GAT. Thus, we feed generated LGRs from ACGAN and SegAN to GAT to obtain the corresponding augmentations for the classification and segmentation tasks, respectively. Hereafter, to differentiate GAN architectures that generate LGR or directly X-ray images, we add -graph at the end to those that generate LGRs. This means that a WGAN that generates LGRs is referred to as a WGAN-graph. 

\begin{figure}[t]
	\centering
	\includegraphics[width=0.96\linewidth]{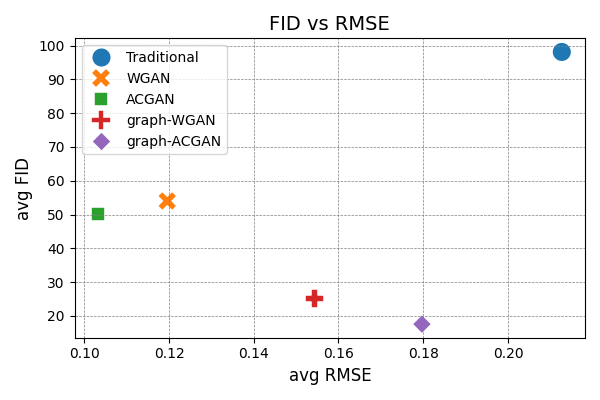}
	\vspace{-10pt}
	\caption{FID vs RMSE: different approaches of data augmentation located in space based on FID and RMSE metrics. 
    Results based on our graph representation produce samples nearer to the distribution (FID $\downarrow$) and with higher distance (RMSE $\uparrow$), demonstrating that we can enhance diversity without compromising the fidelity of the distribution.}
	\label{fig:FIDvsRMSE}
\end{figure}

\begin{figure*}
	\centering
	\includegraphics[width=0.8\linewidth]{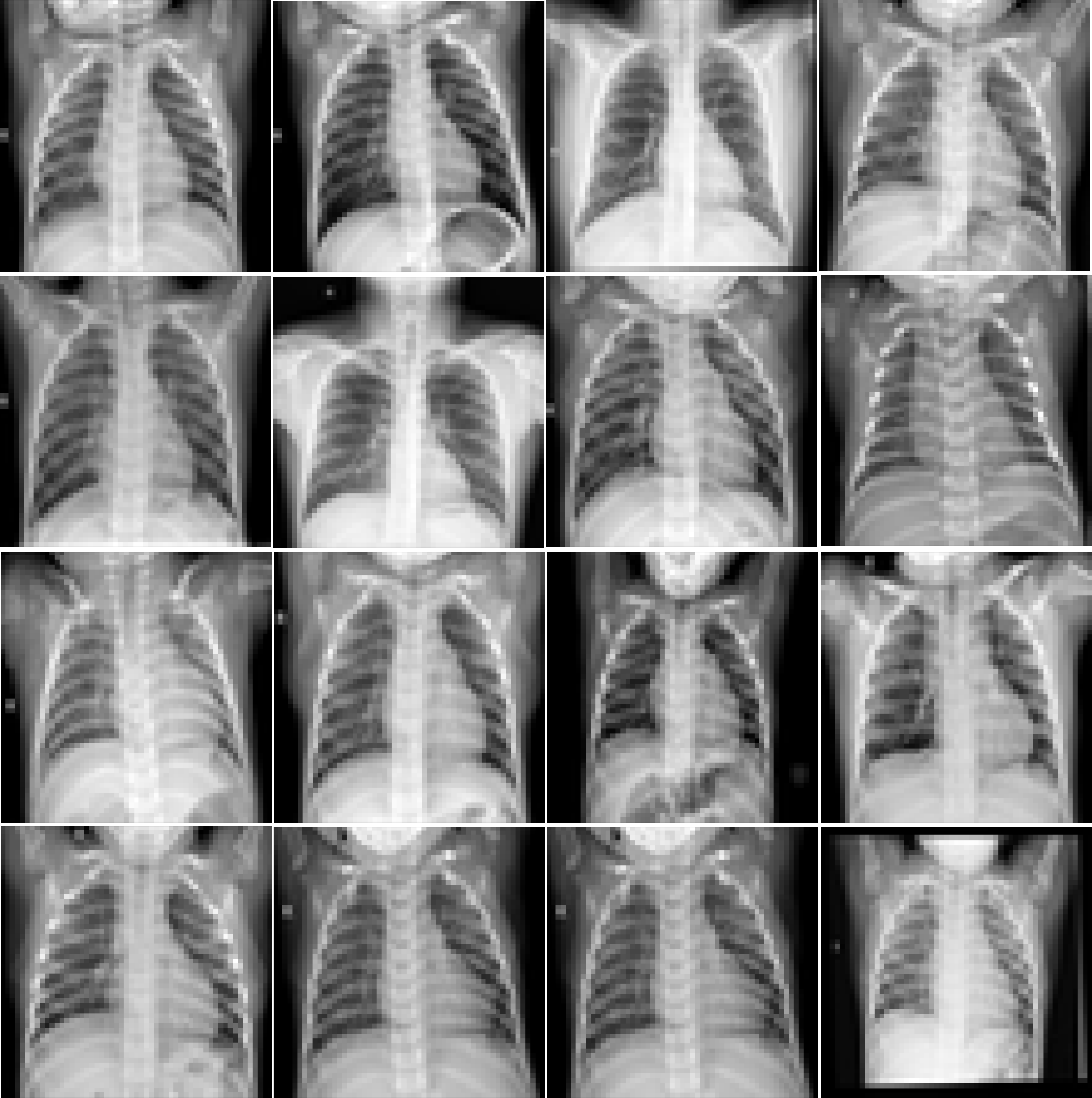}
	\vspace{-10pt}
	\caption{X-ray chest image generation: Image reconstruction is performed from the generated graphs to visually demonstrate the realism and diversity of multiple outputs produced by our LGR. Each image corresponds to a different realization generated by our approach.}
	\label{fig:XrayChestgen}
\end{figure*}

\noindent \textbf{Pneumonia classification}. Our image augmentation pipeline is compared against traditional augmentations and GAN-based augmentations for the classification task. The augmentations based on GANs are inferred from WGAN and ACGAN networks trained on the ``CXR1" dataset. Following the same practice, we \textit{only} employ LGRs from the ``CXR1" dataset to train the graph-WGAN and graph-ACGAN. 
\autoref{tab.cls_aug_comp} shows the accuracy values for different numbers of augmentations $Q=100$, $200$, $500$, and $1000$ of the ``CXR1" dataset for pneumonia classification using the baseline network VGG-16. The FID metric for each augmentation strategy is also presented in \autoref{tab.cls_aug_comp}, showing that our proposed graph-based approaches obtain the best generation quality as well as classification accuracy, and also improve performance while increasing the number of augmentations. 
In order to highlight why our synthesized images are better, in
\autoref{fig:FIDvsRMSE} we show the performance in terms of generation quality (FID) and Euclidean distance (RMSE). We estimate an average FID and RMSE value of $Q=1000$ augmentations for each method. For Traditional WGAN and WGAN-graph methods, the augmentations are independently generated for each class (500 labeled pneumonia and 500 labeled normal), while for ACGAN and ACGAN-graph methods, the augmentations are directly conditioned on the
label (pneumonia/normal) on its architecture. 
We find that our graph-based approaches (WGAN-graph and ACGAN-graph) generate augmentations with more distance from the dataset images ($>$ avg RMSE) than other GAN-based augmentations (WGAN and ACGAN), but closer to the target distribution ($<$ avg FID). 
Therefore, augmentations from our method enable images close to the distribution of chest X-ray images while far from others already existing in the dataset, demonstrating that we can enhance diversity without compromising the fidelity of the distribution. 
Visually, \autoref{fig:XrayChestgen}
displays sample outputs generated by our WGAN-graph approach, which demonstrate the ability of the model to produce realistic and diverse chest X-ray images.

\noindent \textbf{Lung segmentation}.
Likewise, the 'JSRT' dataset is used to train WGAN and SegAN to generate augmentations for lung segmentation. These GANs were adapted to also generate the segmentation mask. \autoref{tab.seg_aug_comp} shows the segmentation performance of the baseline network UNet trained using different numbers of augmentations $Q=0$, $100$, $200$, $500$, and $1000$. The FID metric for each augmentation strategy is also presented in \autoref{tab.seg_aug_comp}. We observe from this table that our augmentation strategy based on the latent graph contributes to creating segmentation examples on training that improve the task performance. Lower FID values for the proposed graph-based generation confirm that the augmented X-ray images are closer to the target distribution.

\begin{table}[t]
	\centering
	\caption{Comparison of classification performance (ACC) of VGG-16 trained using augmented image datasets with traditional X-ray augmentations, GAN-based augmentations, and our graph-based augmentations.}
\resizebox{\columnwidth}{!}{%
	    \begin{tabular}{c|c|c|c|c|c|c}
		\toprule
		\multirow{1}{*}{\small Aug. strategy}  & \multirow{1}{*}{\small FID} &
        \multirow{1}{*}{\small $Q$ $=0$} &
		\multirow{1}{*}{\small $Q$ $=100$} &
		\multirow{1}{*}{\small $Q$ $=300$} &
		\multirow{1}{*}{\small $Q$ $=500$} &
		\multirow{1}{*}{\small $Q$ $=1000$} 
		\\ \hline
        Traditional                
		 & 98.14 & 87.12 & 87.51 & 88.09 & 88.25 & 88.27
		\\ 
		WGAN                 
		 & 54.01 & 87.12 & 88.54 & 89.51 & 90.27 & 90.35
		\\ 
		WGAN-Graph 
		 & 25.13 & 87.12 & \textbf{91.42} & \textbf{92.82} & \textbf{93.25} & \textbf{93.42}  \\ 
        ACGAN 
		 & 50.12 & 87.12 & 89.32 & 90.74 & 92.54 & 92.69  \\  
        ACGAN-Graph
		 & \textbf{17.56} & 87.12 & \textbf{92.24} & \textbf{93.96} & \textbf{94.92} & \textbf{95.12}  \\ \bottomrule
\end{tabular}}
\label{tab.cls_aug_comp}
\vspace{-1em}
\end{table}

\subsection{Representational Power of LGR in GCNs }
Finally, our graph construction is compared against the methods NMGCN \cite{elazab2022novel}, NSCGCN \cite{tang2022nscgcn}, that construct a graph representation using the feature space to then estimate the classification probabilities of X-ray images using GCNs. Thus, the representation ability here is quantified by the impact on the classification accuracy on the CXR2 dataset using SAGPool and having as input the corresponding graph representation. We note that for this experiment, instead of using the self-supervised approach proposed in \eqref{Eq:loss_function1}, we focus on learning the LGR in \eqref{eq:weight_matrix} jointly with the GCN for the classification task. \autoref{tab.cls_comp} shows the comparative results.  
Compared with NMGCN \cite{elazab2022novel}, NSCGCN \cite{tang2022nscgcn}, our graph construction based on learning the topology outperforms all classification performance metrics, F1 score, prediction, accuracy and AUC.\vspace{-1em}

\begin{table}[t]
	\centering
	\caption{Comparison of segmentation performance (DICE) of UNet network trained using augmented image datasets with traditional X-ray augmentations, GAN-based augmentations, and our graph-based X-ray segmentation augmentation. }
\resizebox{\columnwidth}{!}{%
	    \begin{tabular}{c|c|c|c|c|c|c}
		\toprule
		\multirow{1}{*}{\small Aug. strategy} &
		\multirow{1}{*}{\small FID} & 
		\multirow{1}{*}{\small $Q$ $=0$} &
		\multirow{1}{*}{\small $Q$ $=100$} &
		\multirow{1}{*}{\small $Q$ $=300$} &
		\multirow{1}{*}{\small $Q$ $=500$} &
		\multirow{1}{*}{\small $Q$ $=1,000$} 
		\\ \hline
        Trad                
		& 87.72 & 95.40 & 95.76 & 96.07 & 96.22 & 96.21
		\\ 
		WGAN               
		& 53.14 & 95.40 & 96.14 &  96.32 & 96.53 & 96.69
		\\ 
		WGAN-Graph
		& \textbf{29.26} & 95.40 & \textbf{96.16} & \textbf{96.84} & \textbf{97.34} & \textbf{97.55}  \\   
		SegAN                
		& 40.52 & 95.40 & \textbf{96.65}  &  97.26 & 97.54 & 97.57 \\   
		SegAN-Graph                
		& 25.95 & 95.40 & 96.63  & \textbf{97.45} & \textbf{98.05} & \textbf{98.19}
		\\ \bottomrule
\end{tabular}}
\label{tab.seg_aug_comp}
\vspace{-1em}
\end{table}

\begin{table}[t]
	\centering
	\caption{Comparison of classification performance for approaches using latent graph representation NMGCN \cite{elazab2022novel}, NSCGCN \cite{tang2022nscgcn} and ours.}
 \scriptsize
\resizebox{\columnwidth}{!}{%
	    \begin{tabular}{c|c|c|c|c|c}
		\toprule
		\multirow{1}{*}{\small Method}  & \multirow{1}{*}{\small Sen. (\%)} &
        \multirow{1}{*}{\small F1 (\%)} &
		\multirow{1}{*}{\small Pre. (\%)} &
		\multirow{1}{*}{\small Acc. (\%)} &
		\multirow{1}{*}{\small AUC (\%)}
		\\ \hline
		NMGCN \cite{elazab2022novel}
		 & 96.41 & 96.41 & 96.60 & 96.39 & 98.20  \\  
        NSCGCN \cite{tang2022nscgcn}
		 & 96.45 & 96.45 & 96.61 & 96.45 & 99.22  \\ 
        \textbf{Ours}
		 & \textbf{97.52} & \textbf{97.51} & \textbf{97.66} & \textbf{97.52} & \textbf{99.53}   \\ \bottomrule
\end{tabular}}
\label{tab.cls_comp}
\vspace{-3em}
\end{table}

\section{Summary} 
We propose a method for representing chest X-ray images with high structural richness using a latent graph representation (LGR). We propose to optimize our LGR using a novel self-supervised approach jointly with a regularization focusing on important features for medical image diagnosis. We leverage our proposed learned LGR to guide data augmentation in generative models. Experimental results demonstrate that generated LGRs acting as augmentations improve the training performance of GCNs for classification and segmentation tasks. Furthermore, by synthesizing X-ray images from generated LGRs, we demonstrate improved generation results when compared with traditional augmentations or GAN-based augmentations. Our augmentations are nearer to the target distribution while maintaining greater variance, leading to more diverse and effective augmentations.
Our LGR also showed to be competitive for latent representation when compared to other SOTA graph representations for GCNs. 
The flexibility of the proposed approach shows the potential of our LGR to be integrated with more complex networks and extend its applicability to other imaging modalities. 

\bibliographystyle{IEEEtran}
\bibliography{references}

\end{document}